\documentclass[aps,prl,showpacs,twocolumn,latexsym]{revtex4}
\usepackage{amssymb}

\usepackage{latexsym}
\usepackage[dvips]{graphicx}
\usepackage{amsmath}

\begin{document}

\title{ Spin polarization of electrons by non-magnetic heterostructures :
basics of spin-optics. }
\date{\today }
\author{ M.~Khodas,$\,$ A.~Shekhter and A.M.~Finkel'stein }

\begin{abstract}
We propose to use the lateral interface between two regions with different
strengths of the spin-orbit interaction(s) to spin-polarize the electrons in
gated two dimensional semiconductor heterostructures. For a beam with a non
zero angle of incidence the transmitted electrons will split into two spin
polarization components propagating at different angles. We analyze the
refraction at such an interface and outline the basic schemes for filtration
and control of the electron spin.
\end{abstract}

\affiliation{Department of Condensed Matter Physics, the Weizmann Institute of Science,
Rehovot, 76100, Israel}
\maketitle


There is considerable interest in generating spin polarized current in
semiconductor devices for the purposes of spintronics. The central idea is
to polarize electrons using ferromagnetic materials with the subsequent
injection of the polarized electrons into a semiconductor device for further
applications \cite{DattaDas1990}. Despite the noticeable progress in the
understanding the physics of this problem, the technology of the injection
of the spin polarized electrons into a semiconductor system still remains
unsettled; for a review see Ref. \cite{Wolf2001DasSarma2003Ohnol2003}.

In this letter we propose an alternative way to generate a spin polarized
current in heterostructures using non magnetic semiconductor materials only,
see also Ref. \cite{nonmagnetic2002}. We exploit the effect of the
spin-orbit interaction(s) \cite{BychkovRashba84,Dresselhaus} to
polarize the electron beams. The principal element of the proposed spin
polarizer contains an interface between two regions with different strengths
of the spin-orbit interaction(s). As a result of the refraction at such an
interface, for an electron beam with a non zero angle of incidence the
transmitted electrons split into two beams with different spin polarizations
propagating at different angles, and consequently, one can spatially
separate the beams with different polarizations. The further applications of
this effect are similar to that in optical devices exploiting the
polarization of light. The proposed polarizing element can be realized in a
two dimensional (2D) electron (hole) gas confined by an inhomogeneous
quantum well. Such a well can be created either by manipulating the gates 
\cite{Nitta1997,Engels1997,Sato2001,Papadakis2001}, 
or by fabricating a laterally varying heterostructure.
 
Typically, the potential well has the shape of an asymmetric triangle, and,
consequently, there is a direction of asymmetry, $\mathbf{\hat{l}}$,
perpendicular to the electron gas plane. This leads to the appearance of the
Rashba spin-orbit interaction term \cite{BychkovRashba84} in the
Hamiltonian, $\alpha (\mathbf{p\times \hat{l}})\mathbf{\sigma }$. We will
study the case when the parameter $\alpha $ varies along the $x$-direction,
and there is an interface at $x=0.$ The direction of $\mathbf{\hat{l}}$ is
chosen as $\mathbf{\hat{l}}=-\mathbf{\hat{y}}.$ Then the Hamiltonian has the
form: 
\begin{widetext}
\begin{equation}
H_{R} =p_{x}\frac{1}{2m(x)}p_{x}+\frac{1}{2m(x)}p_{z}^{2}+B(x) 
+\frac{1}{2}(\mathbf{\hat{l}\times \sigma })[\alpha (x)\mathbf{p}+\mathbf{p
}\alpha (x)].  \label{eq:Hamiltonian}
\end{equation}
\end{widetext}
Here $B(x)$ describes the varying bottom of the conduction band which may be
controlled by gates. The current operator corresponding to this Hamiltonian
contains a spin-dependent part, $\mathbf{J}=\mathbf{p/}m+\alpha (x)(\mathbf{
\hat{l}\times \sigma })$. The presence of spin in the current operator
implies that in the process of scattering at the interface with varying $
\alpha $ the continuity conditions for the wave function will involve the
spin degrees of freedom of the electrons. The situation is analogous to the
refraction of light where the polarization of light enters the conditions
determining the amplitudes of the refraction (Fresnel formulas).

To diagonalize the Hamiltonian with $\alpha (x)=const$ one has to choose the
axis of the spin quantization along the direction $(\mathbf{\hat{l}\times p}
) $. Then the electron states are described by their chiralities (referred
to as $``+"$ and $``-"$~). For an electron in a state with a definite
chirality the spin polarization is perpendicular to the direction of motion.
The dispersion relations of the two chiral modes are 
\begin{equation}
E^{\pm }=\frac{p^{2}}{2m}\pm \alpha p+B,\quad v=\frac{{\ \partial }E^{\pm }}{
{\ \partial }p}=\frac{p}{m}\pm \alpha .  \label{eq:dispersion}
\end{equation}
Notice that for both modes the velocity depends on the energy in the same
way \cite{Molenkamp2001}, $v=\sqrt{2(E-B)/m+\alpha ^{2}}$ and therefore
under the stationary conditions the two spin components can be separated
only if they are forced to move in different directions \cite{Halperin2003}.

\begin{figure}[h] 
\centerline{ 
    \includegraphics[width=0.45\textwidth]{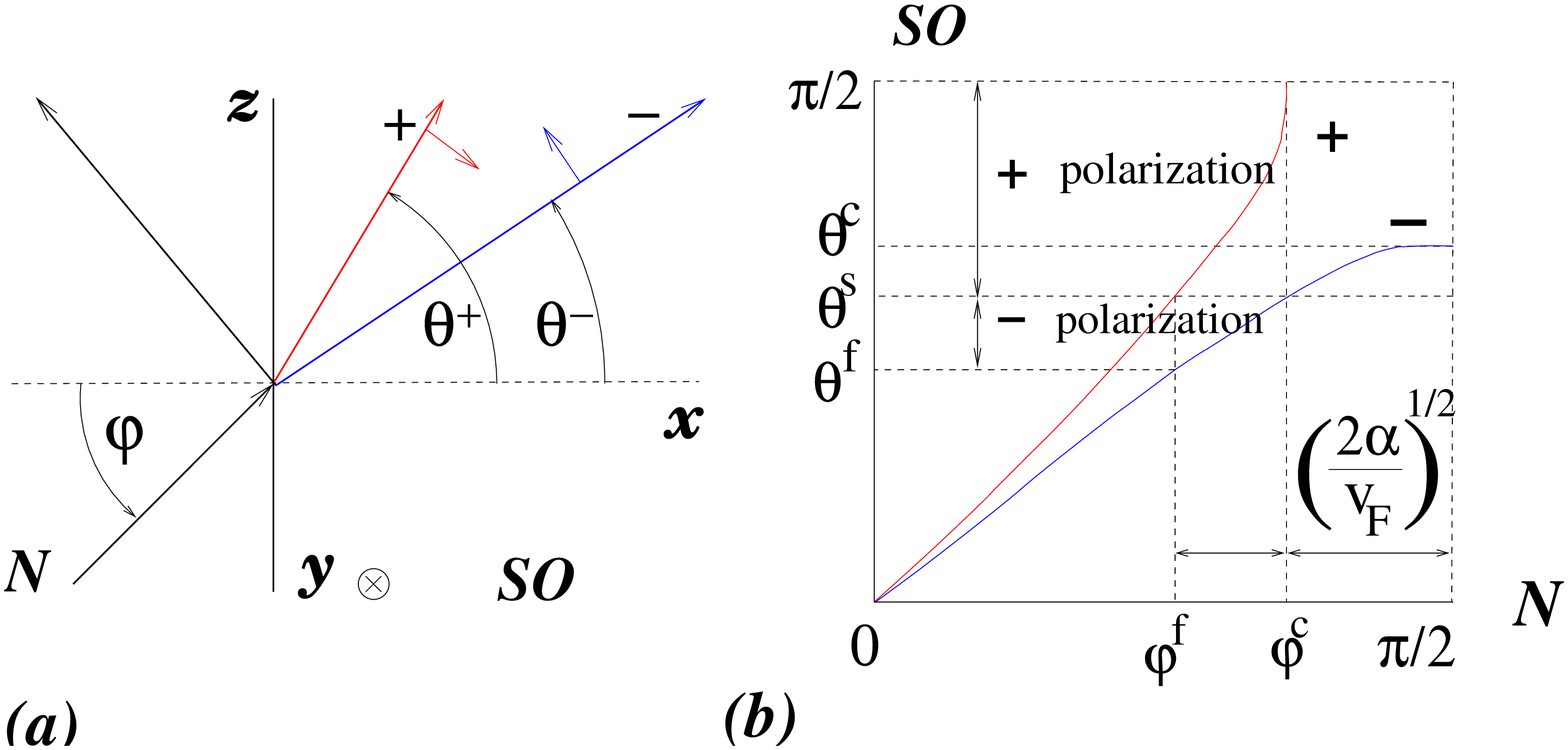}} 
\caption{ (a) The refraction of electrons at the interface between the
regions with (SO) and without (N) spin-orbit interaction. The refracted beam
split into two beams. (b) An angle of refraction for the two polarization
modes as a function of an angle of incidence. $\protect\varphi ^{c}$ is an
angle for total internal reflection for $+$ mode. $\protect\theta ^{c}$
determines the aperture for $-$ mode. Other angles are related to the
spin filtration and are explained in the text. We use $\protect\alpha =0.1$
and $B(x)=const$.}
\label{fig:setup_snell}
\end{figure}

\begin{figure}[t]
\centerline{
\includegraphics[width=0.45\textwidth]{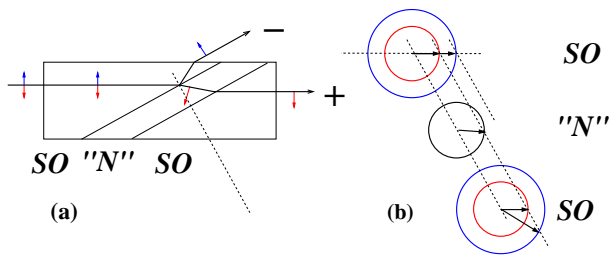} }
\caption{A spin polarizing junction. Two SO-regions are separated by a
stripe with a reduced strength of Bychkov-Rashba term. The stripe leads to
the total reflection of one of the spin components. The kinematical
construction is explained in the text. }
\label{fig:filter}
\end{figure}

\begin{figure}[t]
\centerline{
\includegraphics[width=0.28\textwidth]{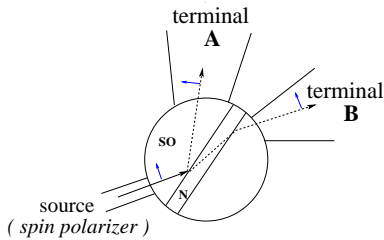} } 
\caption{A spin transistor. A spin current at terminals A and B is
controlled by modulation of the magnitude of $\protect\alpha $. }
\label{fig:spintransistor}
\end{figure}

Let us analyze the kinematical aspects of the scattering at the interface
between the two regions with different $\alpha .$ All the waves
participating in scattering have the same energy $E$ which determines their
momenta as follows: 
\begin{eqnarray}
p^{\pm } &=&m(\sqrt{2(E-B)/m+\alpha ^{2}}\mp \alpha )  \notag \\
&=&mv_{F}(\sqrt{1+\widetilde{\alpha }^{2}}\mp \widetilde{\alpha }).
\label{eq:ppm}
\end{eqnarray}
Here we introduce a small dimensionless parameter 
$\widetilde{\alpha}=\alpha/v_{F}$ 
which we will use throughout the paper. The conservation of
the projection of the momentum on the interface together with Eq.~(\ref
{eq:ppm}) determine the angles of the transmitted and reflected beams
(Snell's law). Figure~\ref{fig:setup_snell} illustrates the scattering for
the simplest case when $\alpha (x<0)=0$. The region without (or suppressed)
spin-orbit term is denoted as N while the region with a finite $\alpha $ is
denoted as SO, and the directions of spin polarizations are indicated by the
small arrows. In Fig. \ref{fig:setup_snell}(a) an incident (unpolarized)
beam comes from the N-region and when transmitted into the SO-region splits
into two beams of different chirality that propagate at different angles.
Thus the interface acts as a spin polarizer. 
In Fig. \ref{fig:setup_snell}(b) the angles of the two beams transmitted
into the SO-region \textit{vs}. the angle of incidence are plotted for the
case $B(x)=const$. From Eq.~(\ref{eq:ppm}) it follows that the SO-region is
optically more dense for the $+$ mode (i.e., it has a smaller wave
vector) and less dense for the $-$ mode. Correspondingly, the $+$ mode
is refracted to larger angles than the $-$ one. Moreover, the $+$ mode
exhibits a total reflection for an angle of incidence in the interval $
\varphi ^{c}<\varphi <\pi /2$ where $\varphi ^{c}$ is a critical angle for
total internal reflection. We will use this fact in the discussion of spin
filtration devices (see Figs. \ref{fig:filter} and \ref{fig:spintransistor}).

Another important fact for the spin filtration is that the $-$ mode has a
limited aperture in the SO-region. Hence, there exists an interval of
outgoing angles, $\pi /2>\theta >\theta ^{c}$, where only the $+$
component can penetrate. If it is possible to collect electrons from this
interval, one will have an ideal spin filter. Potentially promising for spin
filtration is an interval of incident angles $\varphi ^{f}<\varphi <\varphi
^{c}$. For an angle of incidence within this interval the transmitted beams
of different chirality do not overlap. Namely, the $+$ mode scatters into
the interval $\theta ^{s}<\theta <\pi /2,$ while the $-$ mode fills the
interval $\theta ^{f}<\theta <\theta ^{s}$, where $\theta ^{s}$ is the angle
of separation of the two polarizations [see Fig.~\ref{fig:setup_snell}(b)
for a graphical definition of the angles $\varphi^f$, $\theta^f$ 
and~$\theta^s$].

Remarkably, all angle intervals indicated in Fig.~\ref{fig:setup_snell}(b)
are not so narrow as their widths have a square root dependence on $
\widetilde{\alpha }$. It follows from Snell's law that $(\pi /2-\varphi
^{c})\approx (\pi /2-\theta ^{c})\approx \sqrt{2\widetilde{\alpha }}$.
Actually one can reduce $\theta^{c}$ even further. With the gates acting
selectively on the different regions of the electron gas, $\delta
B=B(-\infty )-B(+\infty )\neq 0$, one can alter the position of the bands
relative to the Fermi level in the N- and SO-regions. A simple analysis
based on Eq.~(\ref{eq:ppm}) shows that with an increase of $\delta B$ (i.e.,
lowering $p_{F}$ in the normal region) the angle interval $(\pi /2-\theta
_{c})$ grows and reaches $2\sqrt{\widetilde{\alpha }}$. However, at that
moment, which is optimal for spin filtration, the angle for total internal
reflection reaches $\pi /2$. Starting from this point the angle interval
suitable for spin filtration narrows and eventually becomes $\sim \widetilde{
\alpha }$, instead of $\sim \sqrt{\widetilde{\alpha }}.$

Let us analyze the scattering of electrons at the interface between two
regions with different magnitudes of the Bychkov-Rashba term. The problem
will be considered for the two cases of sharp and smooth interfaces \cite
{Matsuyama2002}. For the clarity of the presentation we limit ourselves to
the case of the interface between the N- and SO-regions only, and it will be
assumed in what follows that $B(x)=const$. The scattering states of an
electron coming from the N-region in the incident state $
e^{i(p_{x}x+p_{z}z)}\chi _{\scriptscriptstyle\mathrm{N}}^{+}$ is given by 
\begin{widetext}
\begin{equation}
\Psi ^{+}=e^{ip_{z}z}\left\{ 
\begin{array}{lllll}
e^{ip_{x}x}\chi _{\scriptscriptstyle\mathrm{N}}^{+}+ & e^{-ip_{x}x}\chi _{
\scriptscriptstyle\mathrm{N}}^{+}r_{++} & + & e^{-ip_{x}x}\chi _{
\scriptscriptstyle\mathrm{N}}^{-}r_{-+}, & x<0 \\ 
& e^{ip_{x}^{+}x}\chi _{\scriptscriptstyle\mathrm{SO}}^{+}t_{++} & + & 
e^{ip_{x}^{-}x}\chi _{\scriptscriptstyle\mathrm{SO}}^{-}t_{-+}, & x>0
\end{array}
\right.  \label{eq:Scattering_State}
\end{equation}
\end{widetext}
where $\chi _{\scriptscriptstyle\mathrm{N/SO}}^{\pm }$ are spinors
corresponding to the $\pm$ chiral modes in the N/SO-regions, and $r$ and 
$t$ are the amplitudes of the reflected and the transmitted waves. A similar
expression holds also for $\Psi ^{-}$ which evolves from the incident state $
\chi _{\scriptscriptstyle\mathrm{N}}^{-}.$

\begin{figure}[h]
\centerline{
    \includegraphics[width=0.3\textwidth]{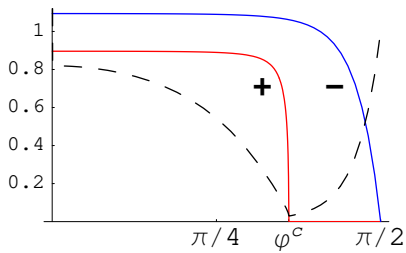}}
\caption{ A sharp N-SO interface; $\protect\alpha=0.1$. The intensities per
unit outgoing angle of the electrons transmitted without change of their
chirality $\sim (d\protect\theta^+/d\protect\varphi)^{-1}\mid t_{++}\mid^2$
and $\sim(d\protect\theta^-/d\protect\varphi)^{-1}\mid t_{--}\mid^2$ as a
function of an angle of incidence. The dashed line indicates the level of
spin polarization, $\mid r_{--} / r_{++} \mid^2 $, of the reflected
electrons for an unpolarized incident beam. }
\label{fig:intensity}
\end{figure}

For the sharp interface the amplitudes $r,t$ can be found from the
continuity conditions that follow from the Schroedinger equation: 
\begin{equation}
\left[ \frac{p_{x}}{m(x)}-\alpha (x)\sigma _{z}\right] \Psi \Big\bracevert 
_{N}^{SO}=0;\quad \quad \Psi \Big\bracevert_{N}^{SO}=0  \label{eq:Derivative}
\end{equation}
where $F\big\bracevert_{N}^{SO}$ denote $F(x=+0)-F(x=-0)$. Analysis of Eq.~(
\ref{eq:Derivative}) shows that in the course of refraction at the interface
with $\widetilde{\alpha }\ll 1$ transitions between waves with different
chiralities are strongly suppressed. Namely, the amplitude $t_{-+}\sim 
\widetilde{\alpha }\left\langle \chi _{\scriptscriptstyle\mathrm{SO}
}^{-}|\chi _{\scriptscriptstyle\mathrm{N}}^{+}\right\rangle \sim \widetilde{
\alpha }^{2}\tan \varphi $, and similarly for $t_{+-}$. An extra factor of $
\widetilde{\alpha }\tan \varphi $ in the off diagonal amplitudes is a
consequence of the fact that angles of deviation of the refracted electrons
are small, and therefore the overlap of the spinors of different chiralities
tends to vanish. The amplitudes $t_{-+}$ and $t_{+-}$ reach their maximal
values $\sim {\widetilde{\alpha }}^{3/2}$ at $\varphi \approx \varphi ^{c}$
where deviation angles are maximal and $\left\langle \chi _{
\scriptscriptstyle\mathrm{SO}}^{-}|\chi _{\scriptscriptstyle\mathrm{N}
}^{+}\right\rangle \sim \sqrt{\widetilde{\alpha }}$. The intensities of the
transmitted electrons without change of their chirality are plotted in Fig. 
\ref{fig:intensity}. The drop of the intensities occurs practically only due
to the reflection which becomes decisive for $\varphi \gtrsim \varphi ^{c}$
. Similar to $t_{+-}$ and $t_{-+}$, the amplitudes of the reflection with a
change of the chirality, $r_{+-}$ and $r_{-+}$, are negligible at any angle.
These amplitudes get their maximal value $\sim \widetilde{\alpha }^{3/2}$ at 
$\varphi =\varphi ^{c}.$ Therefore, when total reflection occurs for
the $+$ mode at $\varphi \geq \varphi _{c}$ its intensity is left mostly in
the same mode.

At angles $\varphi \geq \varphi ^{c}$ the amplitude $r_{++}$ is close to
unity, while $r_{--}$ is still small (as well as $r_{-+}$). It appears that
for the angle of incidence equal to $\varphi _{c}$ the ratio $
|r_{--}/r_{++}|^{2}$ has a cusped minimum. For small $\widetilde{\alpha }$
this ratio has a limiting value $\approx 0.03$ at the minimum. Therefore, an
unpolarized electron beam, when reflected, acquires a significant level of
spin polarization at $\varphi \approx \varphi ^{c}$ (see the dashed line in
Fig. \ref{fig:intensity}). The situation is analogous to the Brewster angle
in the reflection of light. An angular interval around $\varphi _{c}$ where
the degree of polarization of the reflected beam remains large enough is not
so narrow, see Fig. \ref{fig:intensity}. This fact opens an opportunity to
use reflection for the purposes of spin polarization. 

Now we discuss the case of a smooth interface when $\alpha$ changes weakly
on the scale of the electron wavelength $\lambda $. One can conduct the
analysis of the refraction at a smooth interface using a small parameter $
\eta =(d\alpha /dx)/\alpha p_{F}\sim \lambda /d\ll 1,$ where $d$ is a
characteristic scale of the variation of $\alpha $ (i.e., an effective width
of the interface). Due to the smoothness of the interface the electron spin
will adjust itself adiabatically to the momentum keeping its polarization in
the direction perpendicular to the momentum. In addition, for $\eta\ll 1$
the reflected wave can be neglected if $\varphi < \varphi^c$. Having these
arguments in mind, we seek a solution which evolves from the state $\chi
^{+} $ in the form which generalizes the WKB ansatz to include the spin
degrees of freedom: 
\begin{equation}
\Psi ^{+}=\phi _{++}(x)\frac{\chi ^{+}(x)}{\sqrt{v_{x}^{+}}}e^{i\int
p_{x}^{+}{\mathrm{d}}x}+\phi _{-+}(x)\frac{\chi ^{-}(x)}{\sqrt{v_{x}^{-}}}
e^{i\int p_{x}^{-}{\ \mathrm{d}}x}  \label{eq:WKBansatz}
\end{equation}
with $\phi _{++}(x=-\infty )=1$ and $\phi _{-+}(x=-\infty )=0.$

To obtain an admixture of the wave with the opposite chirality, $\phi _{-+}$
and $\phi _{+-}$ $\neq 0$, one has to analyze the Shroedinger equation up to
first order in $\eta $. This equation is similar to the one describing
transitions in a two-level system subjected to an oscillating perturbation
(the Rabi problem \cite{Rabi1937}). The latter arises due to the phase
difference of the two WKB waves in Eq.~(\ref{eq:WKBansatz}). The analysis
shows that the admixture of a wave with different chiralities due to a
smooth interface is very small, $|\phi _{-+}|^{2}\sim \sin ^{2}\varphi (\eta
\alpha _{\scriptscriptstyle\mathrm{SO}}/v_{F})^{2}\ll 1$ or $\sin
^{2}\varphi $($\alpha _{\scriptscriptstyle\mathrm{SO}}/v_{F})^{4}$,
whichever is smaller. In addition, the shape of the $\pm$ lines on Fig. 
\ref{fig:intensity} becomes more rectangular. 

Summarizing the above consideration, one can state that for both the
discussed cases each of the spin chiralities propagates along its own
trajectory, while the change of the chiralities is very inefficient. Hence,
the construction of spin filtering devices should be based on the
kinematical separation of the trajectories of different chiralities. The
N-SO interface analyzed so far for the case of the Rashba spin-orbit
interaction was taken mostly for illustration purposes. Actually, any
lateral interface in the presence of the spin-orbit interaction (of any kind 
\cite{BychkovRashba84,Dresselhaus}) will result in splitting of the
trajectories which can be used for the purpose of spin polarization and
filtration. 

We now consider a spin polarization device presented schematically in Fig. 
\ref{fig:filter}a. The geometry of the device is analogous to the Glan style
optical polarizers made of uniaxial crystals. A stripe with a reduced
strength of the Bychkov-Rashba term is imposed across the SO conductor (SO-$
``$N$"$-SO junction). The direction of the stripe is chosen in such a way
that the angle of incidence of the electron beam exceeds the angle for total
internal reflection for the $-$ mode. (It is the $-$ mode that can be
totally reflected at the SO-$``$N$"$ interface.) The $+$ mode will pass
through the junction mostly unaffected, while the $-$ mode is redirected
as shown in Fig. \ref{fig:filter}a. The reflected $-$ mode carry almost
all of its initial intensity as the change of the chirality is inefficient: $
r_{--}\approx 1$ and $t_{+-},\ r_{+-}\approx 0$. We do not show in Fig. \ref
{fig:filter}a additional beams emerging on each side of the stripe as their
intensity is negligible.

In Fig. \ref{fig:filter}b the kinematics of the refracted electrons is
illustrated. The concentric circles represent spin split Fermi surfaces in
each of the regions of the junction. The dashed lines are directed
perpendicular to the stripe. They show that the projection of momenta on the
direction of the interfaces is conserved. The kinematically allowed wave
vectors in each of the regions are given by the intersection of the dashed
lines with a circle. It is clear from this geometrical construction that for
one of the electron modes to be totally reflected the corresponding dashed
line should not have an intersection with the Fermi surfaces inside the
stripe region. 

The total internal reflection of electrons can be also used as a basis for
the construction of a sort of a spin guide. In a narrow bending stripe of
the ``N''-region tangent electrons in the $+$ state will be trapped
through total internal reflection, while electrons of $-$ chirality will
leak out \cite{nanotech2003}. This guide acts also as a spin polarizer. The
possibility of such device is based on the fact that the intensity of
repolarization $\mid r_{-+}\mid ^{2}$ is very small. 

The high sensitivity of the trajectories of electrons to the magnitude of
the parameter $\alpha $ near the angle for the total internal reflection can
be exploited for the construction of a switch of the spin current (spin
transistor); see Fig. \ref{fig:spintransistor}. Suppose the spin polarized
electron beam is incident on SO-$``$N$"$ (or $``$N$"$-SO) interface with an
the angle of incidence very close to $\varphi _{c}$. Then the spin current
can be switched between the terminals A and B by a small change of $\alpha $
with the gate voltage. In this way an effective modulation of the spin
current can be achieved.

Let us address the question of temperature. The effects under discussion are
mostly controlled by the kinematics. Since the angle for total internal
reflection is different for electrons with different energies, the
temperature leads to a smearing of $\varphi ^{c}$. The polarizing properties
will not be influenced noticeably until the smearing $\delta \varphi
^{c}\lesssim( \pi/2 -\varphi ^{c})$. This leads to the condition: $\delta
\varphi ^{c}/(\pi/2 -\varphi^c ) \approx T/4E_{F}\lesssim 1$. Remarkably,
this condition is not sensitive to the smallness of $\alpha .$

In Ref. \cite{Sato2001} a large spin splitting in a gate controlled electron
gas at In$_{0.75}$Ga$_{0.25}$As/In$_{0.75}$Al$_{0.25}$As\ heterojunction was
reported. The observed splitting corresponds to $\widetilde{\alpha }\approx
0.1$. It was also demonstrated that the parameter $\widetilde{\alpha }$ may
be reduced by a factor of $2$ with the gate voltage. If $\delta \widetilde{
\alpha }_{\scriptscriptstyle\mathrm{SO}}$ across the interface is chosen to
be $0.05$, the interval suitable for spin filtration can be as large as $
26^{\circ }$, which is wide enough for the feasibility of this proposal.

The size of the setup presented in Fig. \ref{fig:spintransistor} is
determined by the distance between the spin polarizer and a stripe
controlling the magnitude $\delta \widetilde{\alpha }_{\scriptscriptstyle 
\mathrm{SO}}$ which can be of the order $\lambda /\cos \theta _{c}\sim
\lambda /\sqrt{\delta \widetilde{\alpha }_{\scriptscriptstyle\mathrm{SO}}}.$
This distance should be shorter than a spin relaxation length.

Finally, we would like to point out the potential advantages of the proposed
method. The spin polarized current can be comparable with the incoming
unpolarized current. The compactness of the proposed setup makes it not very
sensitive to the spin relaxation and disorder. The present experience of
control of ballistic electrons \cite{Stormer2003} makes the proposed method
of spin manipulations feasible.

The authors gratefully acknowledge the discussions with M. Heiblum and Y.
Levinson.

\end{document}